\documentclass[twocolumn,preprintnumbers,amsmath,amssymb]{revtex4-2}
\usepackage{graphicx}  
\usepackage{sidecap}
\begin{document}
%
\title{Micromechanical field-effect transistor terahertz detectors with optical interferometric readout
}
\author{V.~Ryzhii$^{1}$,  C.~Tang$^{1,2}$, T.~Otsuji$^1$, M.~Ryzhii$^{3}$, S. G. Kalenkov$^4$, 
V.~Mitin$^5$, and  M.~S.~Shur$^6$}
\address{
$^1$Research Institute of Electrical Communication,~Tohoku University,~Sendai~ 980-8577, Japan\\ 
$^2$Frontier Research Institute for Interdisciplinary Sciences, Tohoku University, Sendai 980-8578, Japan\\
$^3$Department of Computer Science and Engineering, University of Aizu, Aizu-Wakamatsu 965-8580, Japan\\
$^4$Optoelectronics Department,
Moscow Polytechnic University, Moscow 107023, Russia\\
$^5$Department of Electrical Engineering, University at Buffalo, SUNY, Buffalo, New York 14260 USA\\
$^6$Department of Electrical,~Computer,~and~Systems~Engineering, Rensselaer Polytechnic Institute,~Troy,~New York~12180,~USA\\
}

\begin{abstract} 
\normalsize 
We investigate the response of the micromechanical field-effect transistors (MMFETs) to the impinging
terahertz (THz) signals. The MMFET uses the microcantilevers MC as a mechanically floating gate and the
movable mirror of the Michelson optical interferometer. The MC mechanical oscillations are transformed into
optical signals and the MMFET operates as the detector of THz radiation with the optical output. The
combination of the mechanical and plasmonic resonances in the MMFET with the optical
amplification enables an effective THz detection.
\end{abstract} 
\maketitle

%

%
\newpage
\section{ Introduction}
The concept of a micromechanical field-effect transistor (MMFET) with a 
metal-coated microcatilever (MC) or a
beam serving as a mechanically floating gate was put forward  a long time ago by
Nathanson et al.~\cite{1}. Later, floating-gate MMFETs comprising MCs were fabricated and
measured (see, for instance, also Refs. [2 - 6] and Ref. [7] providing an early review).
MMFETs based on the heterostructures with the two-dimensional electron/hole system
(2DES/2DHS) combine mechanical response with a strong response of the 2DES/2DHS.
They could find applications in micro- and nanoelectromechanical systems (MEMSs and
NEMSs)~\cite{8, 9}. Previously, it was demonstrated that the MC mechanical displacement can
be caused by the ponderomotive force created by the THz radiation input into the MMFET. If
the impinging THz radiation is modulated, the MC motion can become oscillatory. When the
modulation frequency of the THz radiation, $\omega_m$, is close to the frequency,
$\Omega_m$, of the MC
mechanical oscillation, the amplitude of the latter can be resonantly large. The
ponderomotive force induced by the THz radiation can exhibit a substantial increase in the
case of plasmonic resonance (the carrier frequency of THz radiation $\omega$ is close to the
plasmonic frequency, $\Omega$, of the gated 2DES/2DHS in the MMFET). The combination of the
mechanical and plasmonic resonances enables a strong and selective response of the MMFETs as THz detectors~\cite{10,11,12}. This is because of high quality factors of the plasmonic
oscillations (limited by the electron collisions with impurities and phonons~\cite{13,14,15} and by the
electron viscosity~\cite{16}) and very high quality of the MC mechanical oscillations (limited by
different mechanisms of the MC mechanical oscillations damping~\cite{7,9,17,18,19,20}).

In the detectors of the modulated THz radiation on the base of MMFETs considered previously, the variations of the rectified current
were considered as the output signals. However, the mechanical MC oscillations can be detected optically using different interferometric schemes~\cite{21,22,23,24,25,26,27,28}.

In this paper, we propose and analyze  the detectors of modulated THz radiation
based on the MMFETs invoking the optical readout. 
In the detectors under consideration, the MC serves as a movable mirror
for optical radiation in the  Michelson interferometer.

We show that the output optical signal at the modulation frequency can substantially
exceed the direct THz signal response, especially under the conditions of combined
mechanical and plasmonic resonance.

\begin{figure*}[t]
\centering
\includegraphics[width= 16.0cm]{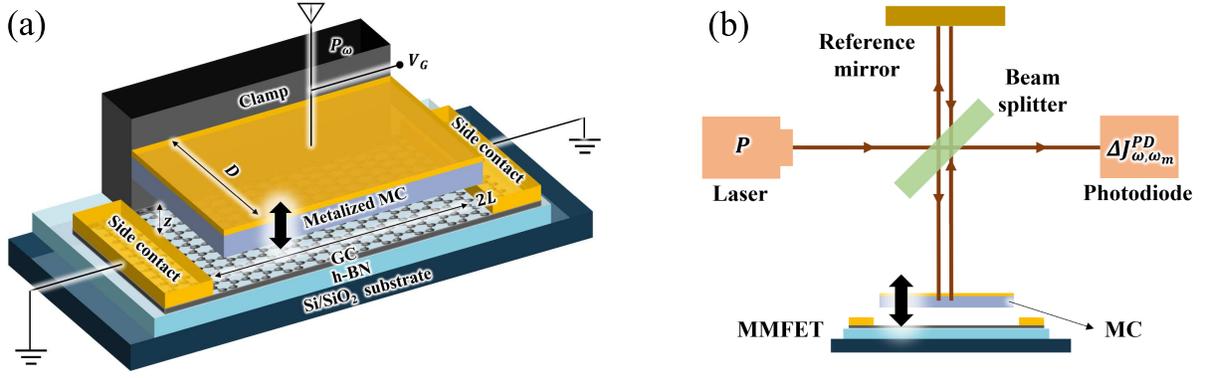}
\caption{Schematic views of (a)  the MMFET  structure and (b) the setup of the MMFET-based  detector scheme with the Michelson interferometer.}
\label{Fig1}
\end{figure*}

\section{MMFET detector structure and  device model}

The MMFET device structure under consideration is schematically shown in Fig. 1(a). We assume that the bias voltage $V_G$ is applied between the clamp-free metal-coated MC, which serves as
the FET mechanically  floating  gate, and the side contacts (source and drain) to the FET channel. For the definiteness, we consider the MMFETs with the MC
with the back clamped  edge, as shown in Fig.~1(a). 
The most interesting properties of the devices under consideration
reveal that when the quality factor of the plasmonic oscillation is sufficiently large,
the electron collision frequency, $\nu$, in the MMFET channel should be a relatively small. Due to this, the MMFETs with the graphene channel (GC) on the h-BN of SiO$_2$ substrate  appear to be rather suitable especially, 
for room-temperature operation.
The results obtained below can also be applied to the MMFETs 
with different methods of their MC fixation at the ends (in particular, to
the MMFETs with clamp-clamp MC), and  
to the MMFETs 
based on more standard heterostructures with the quantum well 2DES/2DHS channel. In the latter case, one needs 
to replace  the GC   fictitious electron/hole mass
by the real effective mass. Our concept with the pertinent modification can also be applied to
 the  nanomechanical FETs (NMFETs).

Despite a relatively complex mechanical movement of the MC with the clamped ends, 
for simplicity, we  consider the MC mechanical properties in
the framework of  the
so-called point-mass model (see, for example, Refs.~[1,\,10,\,29]).
This model is described by 
 the  MC
 point mass
$M$  and the stiffness $K$ of the string
determining its elastic properties.
 The quantities  $M$ and $K$ determine the resonant
frequency of the MC oscillations associated
solely with its mechanical properties, $\Omega_m =\sqrt{K/M}$.
The quantities $M$ and  $\Omega_m$ can be expressed via the MC mechanical parameters: $M = 2LDw\rho$,
$\Omega_m = (w/D^2)\sqrt{E/\rho}$, where $w$, $2L$,and $D$ are the MC thickness,
length, and 
width, and $\rho$ and $E$ are the MC material density and  Young's modulus, respectively.

 The impinging THz radiation with the intensity $I = I_{\omega}[1 + m\sin(\omega_mt)]$,  the carrier  frequency $\omega$, the modulation frequency $\omega_m \ll \omega$, induces the voltage signal between the side contacts
and the MC $\delta V_{\omega}(t) =
\delta V_{\omega} [1 + m\sin(\omega_mt)]\exp(-i\omega t)$, where
$ \delta V_{\omega} = \sqrt{ (A_{\omega} I_{\omega}/gc)}$. Here 
$m \leq 1$ is the modulation depth,
 $A_{\omega}$  and $g$ are the antenna aperture and gain, and $c$ is the speed of light in vacuum. 
 The quantity $(\delta V_{\omega})^2$ can be expressed via the THz power, $P_{\omega}$,
collected by the half-wavelength dipole  antenna~\cite{30,31,32}.  This relationship (in CGS units ) is  $(\delta V_{\omega})^2 = (32P_{\omega}/gc)[1 + m\sin(\omega_mt)]$.
 The bias voltage $V_G$ (gate voltage) allows controlling the MMFET parameters.

The signal voltage results in the oscillations of the channel
potential $\varphi$ and the MC position $Z$.
The variation of the MC position comprises the slowly varying component $\langle Z \rangle$ (which is the MC displacement averaged over the period of the carrier frequency)  and fast oscillating component $\delta\Sigma_{\omega}$: $\Sigma = \langle Z \rangle +\delta\Sigma_{\omega}$. Since the carrier frequency $\omega$ is large (in the THz range) in comparison with the frequency of the MC mechanical oscillations $\Omega_m$ and the modulation frequency $\omega_m$,
$\delta Z_{\omega} \propto \omega^{-2}$ is small. Here and in the following,
the symbol $\langle ...\rangle$ means the averaging over the fast oscillation.
This
approach is similar to the Born-Oppenheimer approximation by introducing two time
scales: fast (to describe the plasmonic oscillations and slow (to describe the MC motion).

Figure 1(b) schematically shows the setup of   THz detector 
incorporating  the optical  Michelson interferometer with the light source (laser),  light-detecting photodiode, and  
the MMFET. The metal coated surface of the MC serves as a movable mirror of the interferometer.
 In
the model under consideration, 
 the averaged (slowly varying) MC mechanical displacement,
$\langle Z\rangle$ is governed
 by the following equation:
\begin{eqnarray}\label{eq1}
\frac{d^2 \langle Z \rangle}{d t^2} + \gamma_m\frac{d\langle Z\rangle}{d t}
+ \Omega_m^2(\langle Z\rangle -W)\nonumber\\
 = -\frac{D}{4\pi\,M}\int_{-L}^{L}dx\langle|{\mathcal E_{\omega}}|^2\rangle,
\end{eqnarray} 
where $W$ is this spacing between the GC and the MC  in the absence
of the applied bias voltage and the THz and optical irradiation, $\gamma_m$ is the damping of the cantilever oscillations associated with different
mechanisms of the energy loss in the cantilever body and
 the clamp.  The axis $z$ is directed perpendicular to the GC and the MC planes,
while the axis $x$ is  along these planes, being directed  from one side contact to the other. 

The term on the right-hand side of Eq. (1)
represents the ponderomotive  force acting on the MC
due  to the applied  bias and the THz signal voltages.
In the situation under consideration,

\begin{eqnarray}\label{eq2}
\langle|{\mathcal E_{\omega}}|^2\rangle  = \frac{(V_G^2 + \langle|\delta \varphi_{\omega}|^2\rangle)}{\langle Z\rangle^2},
\end{eqnarray} 
where $\delta \varphi_{\omega}$ is the signal component of the channel potential oscillating with the frequency $\omega$ (parametrically dependent on the "slow" time).

In the absence of the THz irradiation,  
substituting $\langle|{\mathcal E}|^2\rangle$ from Eq.~(2) into Eq.~(1), and solving the latter, we obtain $\langle Z\rangle \simeq  Z_0$ with  $Z_0$ satisfying the following equation:
\begin{eqnarray}\label{eq3}
Z_0 = W\biggl( 1 
 - \frac{W^2}{Z_0^2}\frac{V_G^2}{{\overline V}_G^2}\biggr).
\end{eqnarray} 
Here
${\overline V_G}  
=\Omega_m\displaystyle \sqrt{4\pi\rho\,w\,W^3} 
=\sqrt{4\pi\,E w^3W^3}/D^2$
is the characteristic gate voltage at which the MC touches the channel, 
$\rho$ is the density of the MC material, and $w$ is the MC thickness. 
At $V_G < {\overline V_G}$, Eq.~(3) yields
\begin{eqnarray}\label{eq4}
 Z_0 \simeq W\biggl(1 - \frac{V_G^2}{{\overline V}_G^2} \biggr).
\end{eqnarray} 
Accounting for Eqs.~(2) and (4), Eq.~(1)  yields for the variation $\Delta Z =\langle Z \rangle - Z_0 $

\begin{eqnarray}\label{eq5}
\frac{d^2  \Delta Z }{d t^2} + \gamma_m\frac{d \Delta Z}{d t}
+ \Omega_m^2 \Delta Z
 = \frac{\overline{\langle|\delta\varphi_{\omega}|^2\rangle}}{4\pi\,\rho\,w\,Z_0^2},
\end{eqnarray} 
where
$\overline{\langle...\rangle} = \displaystyle\frac{1}{2L}\int_{-L}^{L}\langle...\rangle\,dx $ .

\section{Plasmonic ponderomotive force}

Using the  hydrodynamic equations governing the electron transport in the MMFET GC
coupled with the Poisson equation (in the frame of the gradual channel approximation),
we arrive at
\begin{eqnarray}\label{eq6}
\frac{d^2 \delta \varphi_{\omega} }{d x^2} +\alpha_{\omega}^2 \delta \varphi_{\omega}
=0 
\end{eqnarray} 
with the boundary conditions $|\langle\delta \varphi\rangle|_{x = \pm L}|^2 = (\delta V_{\omega})^2[1+m\sin(\omega_mt)]$. 
Here 
\begin{eqnarray}\label{eq7}
\alpha_{\omega}= \frac{\pi\sqrt{\omega(\omega +i\nu)}}{2\Omega}
\end{eqnarray} 
is the plasmonic wavenumber in the gated GC,
\begin{eqnarray}\label{eq8}
\Omega = \frac{\pi\,e}{L\hbar}\sqrt{\mu\,Z_0}  \simeq \Omega_0\sqrt{(1- V_G^2/{\overline V}_G^2)(1+ V_G/{\overline V}_G^*)}
\end{eqnarray} 
is the plasmonic frequency with the electron Fermi energy $\mu \simeq \mu_D$,
where $\Omega_0 = \pi\,e\sqrt{\mu_D\,W}/L\hbar$, $\nu$ is the electron collision frequency in the GC,
$\mu_D \simeq \hbar\,v_W\sqrt{\pi\Sigma_D}$, $\hbar$ is the Planck constant,
$v_W \simeq 10^8$~cm/s, and  $\Sigma_D $ is the donor density in the GC. The characteristic voltage ${\overline V}_G^*$ is calculated in Appendix.

 Solving Eq.~(6) with the above boundary conditions
 we obtain

\begin{eqnarray}\label{eq9}
 |\langle \delta \varphi_{\omega}\rangle|^2 = (\delta V_{\omega})^2 \biggl|\frac{\cos(\alpha_{\omega}x)}{\cos(\alpha_{\omega}L)}\biggr|^2
 [1+m\sin(\omega_mt)].
\end{eqnarray} 
Assuming that $(\Omega/\nu)^2 \gg 1$, we find:

\begin{eqnarray}\label{eq10}
\overline{\langle |\delta \varphi_{\omega}|^2\rangle} \simeq (\delta V_{\omega})^2 
 \frac{[1 +\sin(\pi\omega/\Omega)/(\pi\omega/\Omega)]}
{2|\cos(\pi\sqrt{\omega(\omega+i\nu})/2\Omega)|^2}\nonumber\\ 
\times [1+m\sin(\omega_mt)].
\end{eqnarray} 
Using the representation of trigonometric functions as series~\cite{33},   Eq. (10)
can be presented also in the form

\begin{eqnarray}\label{eq11}
\overline{\langle |\delta \varphi_{\omega}|^2\rangle} \simeq 
(\delta V_{\omega})^2 
 [1 +\sin(\pi\omega/\Omega)/(\pi\omega/\Omega)]\nonumber\\
 \times
\frac{8}{\pi^2}\biggl|\sum_{k=1}\frac{(-1)^{k+1}(2k-1)}
{(2k-1)^2 - \omega(\omega+i\nu)/\Omega^2}\biggr|^2\nonumber\\
 \times[1+m\sin(\omega_mt)].
\end{eqnarray} 
In particular, for the frequency $\omega$ close to the frequency of the fundamental mode of the plasmonic resonance $\Omega$, we obtain

\begin{eqnarray}\label{eq12}
\overline{\langle |\delta \varphi_{\omega}|^2\rangle} 
\simeq (\delta V_{\omega})^2 
\frac{8}{\pi^2}
\frac{[1 +\sin(\pi\omega/\Omega)/(\pi\omega/\Omega)]}{|1 - \omega(\omega+i\nu)/\Omega^2|^2}\nonumber\\
\times[1+m\sin(\omega_mt)].
\end{eqnarray} 

\begin{table*}[t]
\centering
\caption{\label{table} Parameters of the MMFET  detectors with SiC MC ($\rho = 3.2$~g/cm$^{3}$ and $E= 422$~GPa)} 
\vspace{2 mm}
\begin{tabular}{|r|c|c|c|c|c|c|c|c|c|c|c}
\hline
Sample  &$L$($\mu$m)&$D$ ($\mu$m) &$W$($\mu$m)& $w$ (nm)&$\mu$ (meV)&$\Omega/2\pi$ (THz)&$\Omega_m/2\pi$ (MHz)& $Q$ &$Q_m$& ${\overline P}_{\omega}$ \\ 
\hline
MMFET-1&2.0 &4.0	&0.325& 11 &40& 1.29 &1.0&7.3&                                                                                                                                                                                                                                                                                                                                                                                                                                                                                                                                                                                                                                                                                                                                                                                                                                                                                                                                                                                                                                                                                                                                                                                        500 &14\\ 
\hline
MMFET-2&2.0 & 4.0	&0.975& 11 &40& 2.23&1.0&21.8&500                                                                                                                                                                                                                                                                                                                                                                                                                                                                                                                                                                                                                                                                                                                                                                                                                                                                                                                                                                                                                                                                                                                                                                                                                                                                                                                                                                                                                                                                                                                                                                                                                    & 378\\
\hline
MMFET-3&2.0 & 8.0	&0.325& 22 &40& 1.29
 &                                                                                                                                                                                                                                                                                                                                                                                                                                                                                                                                                                                                                                                                                                                                                                                                                                                                                                                                                                                                                                                                                                                                                                                                                                                                                                                                                                                                                                                                                                                                                                                                               0.5&7.3& 500& 7
\\ 
\hline
MMFET-4&2.0 & 4.67	&0.325& 15 &40& 1.29 
&1.0&7.3&315                                                                                                                                                                                                                                                                                                                                                                                                                                                                                                                                                                                                                                                                                                                                                                                                                                                                                                                                                                                                                                                                                                                                                                                                                                                                                                                                                                                                  & 19
\\ 
\hline
MMFET-5&2.0 & 4.0	&0.325& 16.5 &40& 1.29 
&1.5&7.3&220                                                                                                                                                                                                                                                                                                                                                                                                                                                                                                                                                                                                                                                                                                                                                                                                                                                                                                                                                                                                                                                                                                                                                                                                                                                                                                                                                                                                          & 47
\\ 
\hline
MMFET-6&2.0 & 4.0	&0.325& 22 &40&1.29 
&2.0&7.3&62                                                                                                                                                                                                                                                                                                                                                                                                                                                                                                                                                                                                                                                                                                                                                                                                                                                                                                                                                                                                                                                                                                                                                                                                                                                                                                                                                                                                     & 111
\\ 
\hline
\end{tabular}
\end{table*}

 Equations~(5), (9), and (10) yield

\begin{eqnarray}\label{eq13}
 \Delta Z
\simeq \frac{2}{\pi^3}\biggl(\frac{\delta V_{\omega}}{Z_0}\biggr)^2
\frac{ R_{\omega}[1+    mR_{\omega_m}^M\sin(\omega_mt + \theta)]}{\Omega_m^2\,\rho\,w}.
\end{eqnarray} 
Here
\begin{eqnarray}\label{eq14}
 R_{\omega} = \frac{\Omega^4}
{[(\Omega^2 - \omega^2)^2 + \nu^2\omega^2]}
\end{eqnarray} 
and
\begin{eqnarray}\label{eq15}
 R_{\omega_m}^m = \frac{\omega_m^2}
{\sqrt{[(\Omega_m^2 - \omega_m^2)^2 +(\gamma_m\omega_m)^2]}} 
\end{eqnarray} 
are the plasmonic and mechanical  response functions, 
and 
\begin{eqnarray}\label{eq16}
\theta =\tan^{-1}\biggl(\frac{\gamma_m\omega_m}{\omega_m^2-\Omega_m^2}\biggr).
\end{eqnarray} 

Near the combined mechanical and plasmonic resonance ($\omega_m = \Omega_m$ and $\omega = \Omega$), Eqs.~(14) - (16)) yield
\begin{eqnarray}\label{eq17}
R_{\omega_m=\Omega_m}^m \simeq \frac{\Omega_m}{\gamma_m} = Q_m, \qquad 
R_{\omega=\Omega} \simeq \biggl(\frac{\Omega}{\nu}\biggr)^2 = Q,
\end{eqnarray} 
where $Q_m$ and $Q$ are the quality factors of the mechanical and plasmomic resonances,  and 
\begin{eqnarray}\label{eq18}
\theta \simeq \frac{\pi}{2} - \biggl(\frac{\omega_m^2-\Omega_m^2}{\gamma_m\omega_m}\biggr) \simeq \frac{\pi}{2}.
\end{eqnarray} 

Using the relation between the signal voltage $\delta V_{\omega}$ and the THz power $P_{\omega}$ collected by the antenna, from Eqs.~(11)-(15) we obtain
\begin{eqnarray}\label{eq19}
\Delta Z = \Delta Z_{\omega} + \Delta Z_{\omega,\omega_m}
\sin(\omega_m t +\theta).
\end{eqnarray} 
Here
\begin{eqnarray}\label{eq20}
\Delta Z_{\omega}
\simeq 
 \frac{W}{(1-V_G^2/{\overline V}_G^2)^2}
 \biggl(\frac{P_{\omega}}
 {{\overline P_{\omega}}}\biggr)R_{\omega},
\end{eqnarray} 
\begin{eqnarray}\label{eq21}
\Delta Z_{\omega,\omega_m}
\simeq 
 \frac{mW}{(1-V_G^2/{\overline V}_G^2)^2}\biggl(\frac{P_{\omega}}{{\overline P_{\omega}}}\biggr)R_{\omega} R_{\omega_m}^M,
\end{eqnarray} 
and

\begin{eqnarray}\label{eq22}
{\overline P_{\omega}} = 
\biggl(\frac{\pi^3\,g}{64}\biggr)c\rho\,wW^3 \Omega_m^2\propto \frac{w^3W^3}{D^4}
\end{eqnarray}
being the characteristic THz power. 

 Using Eq.~(19),
the amplitude of the MC oscillations at the combined resonance
can be expressed as

\begin{eqnarray}\label{eq23}
\Delta Z_{\omega=\Omega, \omega_m = \Omega_m} \simeq \frac{m W Q_mQ}{(1-V_G^2/{\overline V}_G^2)^2}\biggl(\frac{P_{\Omega}}{{\overline P_{\Omega}}}\biggr).
\end{eqnarray} 
One can see from Eq.~(20) that when the quality factors  of the resonances are high ($Q_m \gg 1$ and $Q \gg 1$) and the modulation depth $m \sim 1$, the MC oscillations amplitude can be comparable with $W$ even at relatively small THz powers ($P_{\Omega} \ll {\overline P_{\Omega}}$).

\section{Optical interferometrical readout}

Considering the device setup with  Michelson interferometer shown in Fig.~1,
for 
the optical power   incident on the optical detector we obtain
\begin{eqnarray}\label{eq24}
P^{PD} = P\biggl\{1+\cos\biggl[\frac{2\pi\,(Z_0 +\Delta Z)}{\lambda}\biggl]\biggr\}.
\end{eqnarray}
Here  $P$ is the  power of light  emitted by the laser and 
$\lambda$ the light wavelength. For brevity, we disregarded  the deviation of the  MC reflection coefficient of light  from unity.

In the following, we assume that $W$ and $V_G$ are chosen to provide
the phase matching $Z_0 = W(1 -V_G^2/{\overline V}_G^2) = 
 (2k-1)\lambda/4$, where $k = 1, 2, 3,...$.
In this case, the  signal component of the optical power received by the photodetector at a relatively weak THz irradiation is equal to 
\begin{eqnarray}\label{eq25}
\Delta P^{PD} = P \sin\biggl[\frac{\pi(2k-1)}{2}\frac{\Delta Z}{W}\biggr].
\end{eqnarray}

Using Eqs.~(19) - (21) and Eq.~(25),  we arrive at 
\begin{eqnarray}\label{eq26}
\Delta P^{PD} = P \sin\biggl[\frac{8W^2}{\pi(2k-1)\lambda^2} \biggl(\frac{P_{\omega}}
 {{\overline P_{\omega}}}\biggr)\nonumber\\
 \times[R_{\omega} + 
 R_{\omega}R_{\omega,\omega_m}m\sin(\omega_mt + \theta)]\biggr].
\end{eqnarray}

Considering that the signal  photocurrent produced by the photodiode 
$\Delta J^{PD} = e\eta\,\Delta P^{PD}\lambda/\pi\hbar\,c$, where
$\eta$ is the photodiode quantum efficiency and $e$ is the electron charge,
 and 
using $ \Delta P^{PD}$ given by Eq.~(26), we obtain
 for  relatively weak THz powers
\begin{eqnarray}\label{eq27}
\Delta J_{\omega}^{PD} = \Delta J_{\omega}^{PD} + \Delta J_{\omega,\omega_m}^{PD}\sin(\omega_mt + \theta),
\end{eqnarray} 
where
\begin{eqnarray}\label{eq28}
\Delta J_{\omega}^{PD} \simeq
\biggl(\frac{8e\eta\,W^2}{\pi^2(2k-1)\hbar\,c\lambda}\biggr)\biggl(\frac{PP_{\omega}}{{\overline P_{\omega}}}\biggr) R_{\omega}
\end{eqnarray} 
and

\begin{eqnarray}\label{eq29}
\Delta J_{\omega,\omega_m}^{PD} \simeq
m\biggl(\frac{8e\eta\,W^2}{\pi^2(2k-1)\hbar\,c\lambda}\biggr)\biggl(\frac{PP_{\omega}}{{\overline P_{\omega}}}\biggr) R_{\omega}R_{\omega,\omega_m}
\end{eqnarray} 
are the steady-state (rectified) and modulated components of the output current.

\begin{figure}[t]
\centering
\includegraphics[width= 9.0cm]{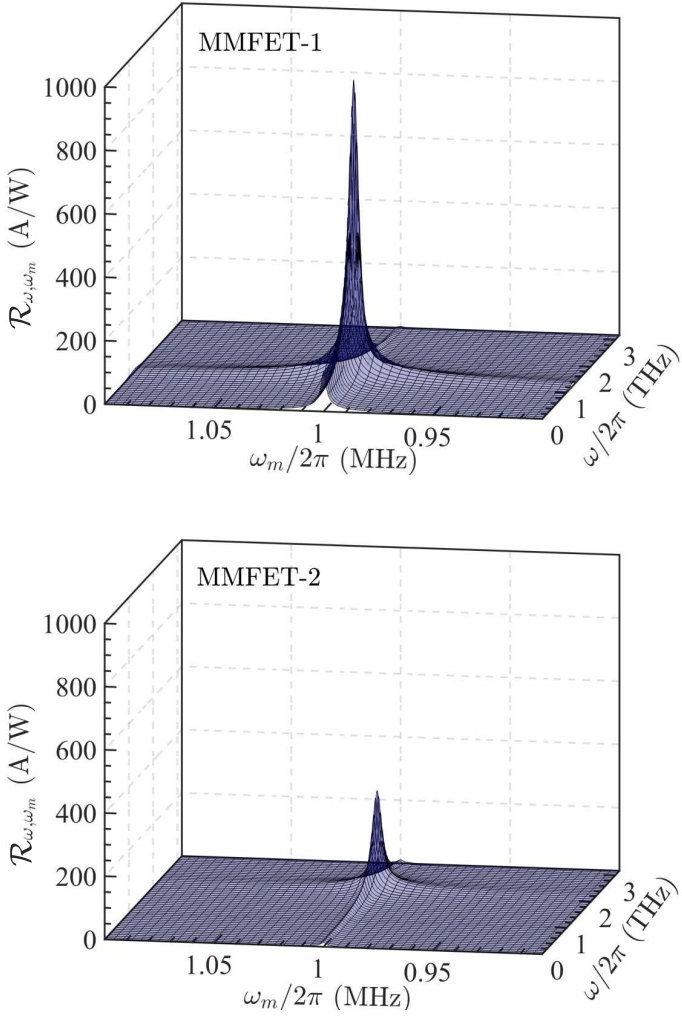}
\caption{
The responsivity ${\mathcal R}_{\omega,\omega_m}$  of the MMFET detector versus the carrier, $\omega/2\pi$, and modulation,  $\omega_m/2\pi$, frequencies, 
for   MMFET-1: $\Omega/2\pi =1.29$~THz, $\Omega_m/2\pi = 1.0$~MHz (upper panel) and (for   MMFET-2: $\Omega/2\pi =2.23$~THz, $\Omega_m/2\pi = 1.0$~MHz (lower panel).}
\label{Fig2}
\end{figure}

\section{MMFET  detector responsivity}

For the MMFET THz detector responsivity, defined as 
${\mathcal R}_{\omega,\omega_m} =  \Delta J^{PD}_{\omega,\omega_m}/P_{\omega}$,  Eq.~(29) yields

\begin{eqnarray}\label{eq30}
{\mathcal R}_{\omega,\omega_m} \simeq m {\mathcal R}
\biggl(\frac{P}{{\overline P_{\omega}}}\biggr)
R_{\omega} R_{\omega_m}^m
\end{eqnarray} 
with the responsivity characteristic value ${\overline {\mathcal R}}$ given by 
\begin{eqnarray}\label{eq31}
 {\mathcal R} = \frac{8e\eta\,W^2}
{\pi^2(2k-1)\hbar\,c\lambda}.
\end{eqnarray} 

 Assuming here and in the following (for definiteness) that the "optical" parameters are  $\lambda = 1.3~\mu$m,  $\eta = 1$,  and the laser power $P= 10$~mV for  MMFET-1 and MMFET-2 with
the values  $W$ corresponding to the chosen laser wavelength (at $V_G = 0$), from Eq.~(31) we obtain ${\mathcal R} \simeq (0.35 - 1.05)$~A/W.
 
Figures~2 and 3 show  examples of the responsivity, ${\mathcal R}_{\omega,\omega_m}$, as a function of the carrier, $\omega/2\pi$, and modulation,
$\omega_m/2\pi$ frequencies calculated using Eqs.~(30) and (31) and invoking Eqs.~(14) and (15). The MMFET parameters used in the calculations are listed in Table I.
It is assumed that the bias voltage $V_G = 0$ and the  electron collision frequency 
$\nu = 3$~ps$^{-1}$ (that  corresponds to the 2DES mobility in the GC $\mathcal M \simeq 3\times 10^4$~cm$^2$/Vs). 
The mechanical parameters correspond to the  metal coated
SiC MCs.
Considering the MMFETs with different parameters $D$ and $w$, we account for that the quality factor of the mechanical resonance  $Q_m \propto (D/w)^3$ (see, for example,~Ref.~[34]).

 Figure~2 shows  the responsivity ${\mathcal R}_{\omega,\omega_m}$
of MMFET-1 and MMFET-2 
as a function of $\omega$ and $\omega_m$. It is seen that the responsivity exhibits  large peak values (at the combined resonance) and fairly pronounced spectral selectively, especially in respect to the modulation frequency.
The comparison of the MMFET-1 and MMFET-2 responsivities shows that an increase
in $W$ leads to a marked decrease in the peak height. This is because of the dependence of the characteristic THz power $\overline P_{\omega}$ on  $W$
given by Eq.~(22). 

\begin{figure}[t]
\centering
\includegraphics[width= 9.0cm]{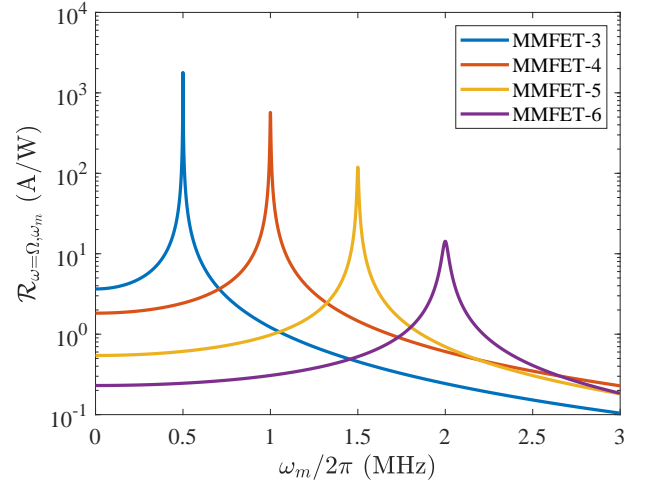}
\caption{
The responsivity ${\mathcal R}_{\omega,\omega_m}$ of the detectors with different parameters calculated as a function of  the  modulation frequency   $\omega_m/2\pi$  at the plasmonic resonance ($\omega/2\pi = \Omega/2\pi = 1.29$~THz).}
\label{Fig3}
\end{figure}

Figure~3 shows the responsivity ${\mathcal R}_{\Omega,\omega_m}$
of MMFET-3 -  MMFET-6  versus the modulation
frequency $\omega_m/2\pi$ [accounting for that the  mechanical quality factors $Q_m \propto (D/w)^3$] and  assuming that the carrier frequencies are equal to the  plasmonic resonant frequency (i.e., $\Omega/2\pi = 1.29$~THz).
One can see that the resonant peak  responsivity is
 fairly high and sharp, and their height markedly drops with increasing
MC thickness $w$. This is because an increase in $w$ reinforces the MC mechanical stiffness and, hence, leads to  an increase in  the mechanical resonance frequency $\Omega_m \propto w$ and increase in the characteristic  THz power ${\overline P}_{\omega } \propto w^3$ [see Eq.~(22) and Table I].

\section{Comments}
\subsection*{Voltage tuning}
\vspace{-3mm}
 In the above demonstration of the spectral characteristics of the MMFET detector responsivity, we assumed that the spacing $W$ (or the laser radiation wavelength $\lambda$) is chosen to correspond
to the interferometer resonance at $V_G =0$ ($W = \lambda/4$ or $W = 3\lambda/4$). If this condition is not fulfilled, for example, $W$ somewhat exceeds $\lambda/4$ or $3\lambda/4$, the bias voltage can be used to tune the spacing between the GC and the MC to provide a proper phase difference of the light beams in the interferometer and achieve  phase matching.
This is owing to
the variation of the voltage $V_G$ varies the MC position $Z_0$.

Let us consider the case when $4W/(2k-1)\lambda = 1 + \delta$, where $\delta \ll 1$.
In such a case, accounting for the voltage dependence of $Z_0$ given by Eq.~(4), the phase resonance $4Z_0/(2k-1)\lambda = 1$ is achieved if $V_G^2/{\overline V}_G^2 = \delta$. 
Using the date listed in Table I, the characteristic voltages for MMFET-1
and MMFET-2 are
 about ${\overline V}_G \simeq2.3$~V and ${\overline V}_G \simeq 12.0$~V, respectively. Hence, 
at $\delta = 5~\%$, taking into account Table I, the tuning voltage is $V_G \simeq (0.52 - 2.68)$~V.

The variation of the bias voltage affects the plasmonic frequency $\Omega$.
This can result in a detuning of the plasmonic resonance.

 The plasmonic frequency depends on $V_G$ via the dependence on the spacing between the MC and GC and via the dependence on the Fermi energy: $\mu \simeq \mu_D\sqrt{1+ \displaystyle\frac{\mu_0eV_G}{\mu_D(\mu_0+\mu_D)}} \simeq \mu_D\biggl(1 - \displaystyle\frac{V_G}{{\tilde V}_G}\biggr)$ (see, for example, Ref.~[32]). In the latter formula, $\mu_0 =\hbar^2v_W^2/4e^2W$ and ${\tilde V}_G = 2\mu_D(\mu_D+\mu_0)/e\mu_0$. For $\mu = 40$~meV and $W = (0.325 - 0.975)~\mu$m, we find ${\tilde V}_G
 \simeq (15.39 - 45.91)$~V.  
 At moderate bias voltages, Eq.~(8) yields
$\displaystyle  \frac{\Omega-\Omega_0}{\Omega_0} \simeq \frac{1}{2}\biggl(\frac{V_G}{{\tilde V_G}} - \frac{V_G^2}{{\overline V}_G^2}\biggr)$.
 The voltage detuning of the plasmonic resonance might be crucial
 if $(\Omega-\Omega_0)/\Omega_0 \gtrsim Q^{-1}$. 
 Using the latter relations, assuming that $\delta = 5~\%$, and   accounting for  the above estimates for ${\overline V}_G$ and ${\tilde V}_G$,
 we find $(\Omega-\Omega_0)/\Omega_0 \lesssim 0.01$, i.e., much smaller than $Q^{-1} =(\nu/\Omega)^2$. Hence,   the detuning
 of the plasmon resonances in the MMFETs under consideration can be disregarded. 

The voltage tuning of the mechanical resonances is, in principle, also  possible.
However, such  tuning is associated with the nonlinearity of the MC oscillations (see, for example, Refs.~[18,35].
This effect is fairly weak, at least in the devices under consideration and therefore was disregarded in our model.

\subsection*{Effect of the passivation layer}
\vspace{-3mm}
 The GC can be covered by the passivation layer. Its thickness $h$ should be smaller than the thickness of the air (or vaccum)  $h \ll W - w$ not to prevent the MC free oscillations.
If the dielectric constant of the passivation layer is relatively high
(for example, it is made of HfO$_2$), this layer can reinforce the ponderomotive force acting on the MC and somewhat decrease the plasmonic frequency.

\subsection*{Light pressure}
\vspace{-3mm}
 We neglected the MC displacement caused by the pressure of the light emitted by the interferometer laser incident on and reflected from the MC.
The force acting on the MC is equal  to $F =2P/c$.
This force leads to the MC displacement equal to $\Delta Z^{OPT} =2P/cM\Omega_m^2$. 
Such   effect is insignificant  if $\Delta Z^{OPT} \ll W$. The latter condition leads to   $P \ll {\overline P} =
M\Omega_M^2cW/2$. For the MMFETs with the parameters indicated in Table I,
 $M= 2LDw\rho \simeq (5 - 15)\times 10^{-13}$~g, we obtain ${\overline P} \simeq (0.96-14.4)$~W. This implies that for $P$ in the mW range, $\Delta Z^{OPT} \ll W$. In this power range,  neglecting of the light pressure effect is justified.
The ratio ${\overline P}_{\omega}/{\overline P} =(\pi^3g/32)( W^2/LD) \ll 0.025 - 0.075$.

\subsection*{Comparison with the FET THz hot-electron  bolometers}
\vspace{-3mm}
The FET-based  hot-electron bolometers, considered in Refs.~[31,32], can also   detect
 the modulated THz. This is because their bandwidth is fairly broad, substantially covering the MHz range of the modulation frequencies.
The main distinction between the MMFET-based detector analyzed above and
the FET-based bolometers is the extremely strong frequency selectivity of the former (near the combined mechanical and plasmonic resonance).  
As can be shown, the current responsivities of these two types of THz photodetectors  are rather close to each other. However, one needs to note that
the resistance of the photodiode producing the output current can be very
high (in the M$\Omega$ range for InGaAs photodiodes). This implies
that the MMFET-based detectors can exhibit much higher voltage responsivities.
 
\subsection*{Operation at elevated modulation frequencies} 
\vspace{-3mm}
An increase in the mechanical resonance frequency can be achieved using
the MCs with much smaller width $D$ (see, for example, Ref.~[9])  separated by  narrower spacing $W$.
Indeed, assuming that $D = 1~\mu$m, $W=0.05~\mu$m (50~nm),  and $w = 11$~nm,
we obtain $\Omega_m/2\pi = 16$~MHz with the  characteristic power
${\overline P}_{\omega}$ of the same order of magnitude as shown in Table I for MMFET-1.
However, a decrease in the MC width $D$ might necessitate  shortening of the GC length $2L$. The latter can lead to the excessive plasmonic resonance frequency
$\Omega$. As a result, the carrier frequency $\omega$ can be much smaller than $\Omega$, i.e., far from the plasmonic resonance, negatively affecting
the detector responsivity.   Another issue is that a decrease in $D$ and $L$
is limited by the laser radiation wavelength $\lambda$ due to the light diffraction on the narrow MC~\cite{23}. The consideration 
of the MMFET-based  detectors of the THz radiation modulated at elevated
modulation frequencies requires a separate study.

\section*{Conclusions}

We proposed the detectors of the modulated THz radiation based on the MMFETs
using the optical interferometric readout. These detectors use the combined 
mechanical and plasmonic resonances,  achieving a strong response 
when the modulation and carrier frequencies are close to the frequency of the MC mechanical oscillation and the plasmonic frequency, respectively.
Due to a high quality factor of the mechanical resonance, the MMFET-based detector can exhibit a strong selectivity with respect to the modulation frequency. The MMFET-based detectors can be used in the free-space THz communications and other THz applications.

\section*{Author's contributions}
\vspace{-3mm}
All authors contributed equally to this work.
\vspace{-3mm}
\section*{Acknowledgments}
\vspace{-3mm}
V.~R. and S.~K. are grateful to Prof. V. G. Leiman for stimulating discussions.
The work  was supported by the Japan Society for Promotion of Science (KAKENHI  Nos. 21H04546, 20K20349),
Japan, the RIEC Nation-Wide Collaborative research
Project No. R04/A10, Japan, and  by AFOSR (contract number FA9550-19-1-0355).

\section*{Conflict of Interest}
\vspace{-3mm}
The authors declare no conflict of interest.

\section*{Data availability}
\vspace{-3mm}
All data that support the findings of this study
are available within the article.


\begin{thebibliography}{1}
\bibitem{1} 
H. C. Nathanson, W. E. Newell, R. A. Wickstrom, and
J. R. Davis, Jr.,\lq\lq The resonant gate transistor,\rq\rq IEEE Trans. Electron Devices, {\bf 14}, 117
(1967).

\bibitem{2} 
M. Roukes,
\lq\lq Nanoelectromechanical systems face the future, \rq\rq\, Physics World
{\bf 14}, 25 (2001).)

\bibitem{3}  
W. H. Teh, R. Crook, C. G. Smith, H. E. Beere, and
D. A. Ritchie,
\lq\lq Characteristics of a micromachined floating-gate high-electron-mobility 
transistor at 4.2 K,\rq\rq\,  J. Appl. Phys. {\bf 97}, 114507 (2005).

\bibitem{4} 
 R. G. Beck, M. A. Eriksson, R. A. Westervelt,
K. L. Campman, and A. C. Gossard,
\lq\lq Strain‐sensing cryogenic field‐effect transistor for integrated strain detection in GaAs/AlGaAs microelectromechanical systems,\rq\rq\,
 Appl. Phys Lett.
{\bf 68}, 3763 (1996).



\bibitem{5} 
 M. P. Schwarz, D. Grundler, I. Meinel, Ch. Heyn, and
D. Heitmann, \lq\lq Micromechanical cantilever magnetometer with an integrated two-dimensional electron system,\rq\rq\, Appl. Phys Lett. 76, 3464 (2000).

\bibitem{6}  H. Yamaguchi, S. Miyashita, and Y. Hirayama,
\lq\lq Microelectromechanical displacement sensing using InAs/AlGaSb heterostructures,\rq\rq\,
Appl. Phys Lett. {\bf 82}, 394 (2003).


\bibitem{7} 
 K. L. Ekunchi and M. L. Roukes,
\lq\lq Nanoelectromechanical systems,\rq\rq\, Rev. Sci. Inst. {\bf 76},
161101 (2005).





\bibitem{8} 
Y. Tsuchiya, K. Takai, N. Momo, T. Nagami, H. Mizuta,
S. Oda, S. Yamaguchi, and T. Shimada, \lq\lq Nanoelectromechanical nonvolatile memory device incorporating nanocrystalline Si dots,\rq\rq J. Appl. Phys.
{\bf 100}, 094306 (2006).


\bibitem{9} 
Bo Xu,  P. Zhang, J. Zhu, et al.
\lq\lq Nanomechanical resonators: Toward atomic
scale,\rq\rq ACS Nano {\bf 16}, 15545 (2022).


\bibitem{10} 
V. Ryzhii, M. Ryzhii, Y. Hu, I. Hagiwara, and M. S. Shur
\lq\lq Resonant detection of modulated terahertz radiation in micromachined high-electron-mobility transistor,\rq\rq
Appl. Phys. Lett. {\bf 90}, 203503 (2007).


\bibitem{11} 
Y. Hu, M. Ryzhii, I. Hagiwara, M. S. Shur, and V. Ryzhii, \lq\lq Combined resonance and resonant detection of modulated terahertz radiation in a micromachined high‐electron mobility transistor,\rq\rq Phys. Stat. Sol. C
{\bf  5}, 277 (2008).

\bibitem{12} 
V. G. Leiman, M. Ryzhii, A. Satou, N. Ryabova, V. Ryzhii, T. Otsuji, and M. S. Shur,
\lq\lq Analysis of resonant detection of terahertz
radiation in high-electron mobility transistor
with a nanostring/carbon nanotube as
mechanically floating gate,\rq\rq
J. Appl. Phys. {\bf 104}, 024514 (2008).


\bibitem{13}  M. Dyakonov and M. Shur,\lq\lq Plasma wave electronics: novel terahertz devices using two dimensional electron fluid,\rq\rq\, IEEE Trans. Electron Devices
{\bf 43}, 1640 (1996).

\bibitem{14} 
V. Ryzhii, A. Satou, and T. Otsuji,
\lq\lq Plasma waves in two-dimensional electron-hole system in gated graphene heterostructures,\rq\rq\,
J. Appl. Phys. {\bf 101}, 024509 (2007).



\bibitem{15} 
 V. Ryzhii, T. Otsuji, and M. S. Shur,
 \lq\lq Graphene based plasma-wave devices for terahertz applications,\rq\rq\,
 Appl. Phys. Lett. {\bf 116}, 140501 (2019).

\bibitem{16}
Y. Zhang and M. S. Shur, \lq\lq Collision dominated, ballistic,
and viscous regimes of terahertz plasmonic detection by
graphene,\rq\rq\, J. Appl. Phys. {\bf 129}, 053102 (2021).




\bibitem{17}
Y. Xu, C. Chen, V. V. Deshpande, F. A. DiRenno,
A. Gondarenko, D. B. Heinz, S. Liu, P. Kim, and J. Hone,
\lq\lq Radio frequency electrical transduction of graphene mechanical resonators,\rq\rq\, Appl. Phys. Lett. {\bf 97}, 243111 (2010). 


\bibitem{18}
A. Eichler, J. Moser, J. Chaste, M. Zdrojek, I. Wilson-Rae, and A. Bachtold,
\lq\lq Nonlinear damping in mechanical resonators made
from carbon nanotubes and graphene,\rq\rq\,
Nat. Nanotech. {\bf 6}, 339-342 (2011).

\bibitem{19}
N. Morell, A. Reserbat-Plantey, I. Tsioutsios, K G. Scheadler, F. Dubin, F. H. L. Koppens, and A. Bachtold, \lq\lq High quality factor mechanical resonators based on WSe$_2$ monolayers,\rq\rq\,
Nano Lett, {\bf 16} 5102 (2016)




  \bibitem{20}
  G. Aoust, R. Levya, B. Bourgeteaua, and O. Le Traon,
\lq\lq Viscous damping on flexural mechanical resonators,\rq\rq
Sens. and Actuators A Phys. {\bf 230}, 126 (2015) 
 


\bibitem{21} 
D. Karabacak, T. Kouh, and  K. L. Ekinci,
\lq\lq Analysis of optical interferometric displacement detection in nanoelectromechanical systems,\rq\rq\,
J. Appl. Phys. {\bf 98}, 124309 (2005).


\bibitem{22} 
L. A. J. Davis, D. R. Billson, D. A. Hutchins, and  R. A. Noble.
\lq\lq 
Visualizing acoustic displacements of capacitive micromachined transducers using an interferometric microscope,\rq\rq,
Acoustics Res. Lett. Online {\bf 6}, 75–79 (2005).

\bibitem{23}
T. Kouh, D. Karabacak, D. H. Kim, and K. L. Ekinci,
\lq\lq Diffraction effects in optical interferometric displacement detection in
nanoelectromechanical systems,\rq\rq\,
Appl. Phys. Lett.  {\bf 86}, 013106 (2005). %

\bibitem{24} 
J. Wehrmeister, 
A. Fuss, F. Saurenbach, R. Berger, and M. Helm,
\lq\lq Readout of micromechanical cantilever sensor arrays by Fabry-Perot
interferometry,\rq\rq\,
Rev. Sci. Instrum. {\bf 78}, 104105 (2007).

\bibitem{25} 
Z.  Wang and P. X-L Feng, \lq\lq  Interferometric motion detection in atomic layer 2D nanostructures: visualizing signal transduction
efficiency and optimization pathways,\rq\rq Sci. Rep. {\bf 6} 28923 (2016).

\bibitem{26} R. J. Dolleman, D. Davidovikj, 
 H. S. J. van der Zant, and P. G. Steeneken,
\lq\lq Amplitude calibration of 2D mechanical resonators by nonlinear optical transduction,\rq\rq
Appl. Phys. Lett. {\bf 111}, 253104 (2017).

\bibitem{27} 
R. de Alba,  C. B. Wallin, G. Holland, S. Krylov, and B. R. Ilic,
\lq\lq Absolute deflection measurements in a micro- and nano-electromechanical Fabry-Perot interferometry system,\rq\rq\, J. Appl. Phys. {\bf 126}, 014502
(2019).



\bibitem{28}
J. Zhu, P. Zhang, R. Yang, and  Z. Wang,
\lq\lq Analyzing electrostatic modulation of signal
transduction efficiency in MoS$_2$
nanoelectromechanical resonators with
interferometric readout,\rq\rq\,
Sci. China, Inf. Sci. {\bf 65}, 122409: 1-7 (2022). 

 
\bibitem{29} 
 J.-S. Wu and  T.-L. Lin,
 \lq\lq Free vibration analysis of a uniform cantilever beam with point masses by an analytical-and-numerical-combined method,\rq\rq\,
 J. Sound. Vib. {\bf 136}, 201 (1990).
 
 
 
 
\bibitem{30}  
R. E.  Colin,{\it Antenna and Radiowave Propagation}(New York, McGraw-Hill, 1985).

\bibitem{31}
V. Ryzhii, C. Tang, T. Otsuji, M. Ryzhii, V. Mitin, and M. S. Shur, \lq\lq Effect of electron thermal
conductivity on resonant plasmonic detection in the
metal/black-AsP/graphene FET terahertz hot-electron
bolometers,\rq\rq Phys. Rev. Appl. {\bf 19}, 064033 (2023).


\bibitem{32}
V. Ryzhii, C. Tang, T. Otsuji, M. Ryzhii, V. Mitin, and M. S. Shur,
\lq\lq Hot-electron resonant terahertz bolometric detection in the graphene/black-AsP
field-effect transistors with a floating gate,\rq\rq\,
J. Appl. Phys. {\bf 133}, 174501 (2023).

\bibitem{33}
 I. S. Gradstein and I. M. Ryzhik, {\it Tables of Integrals,
Series, and Products}, Academic Press, New York 1994,
p. 43.
 
 \bibitem{34}
Zh. Hao, A. Erbil, and  F. Ayazi, \lq\lq An analytical model for support loss in micromachined
beam resonators with in-plane flexural vibrations,\rq\rq
Sens.  Actuator A Phis. {\bf 109}, 156 (2003). 

\bibitem{35}
Q. P. Unterreithmeier, T. Faust, and J. P. Kotthauss, \lq\lq
Nonlinear switching dynamics
in a nanomechanical resonator,\rq\rq\, Phys. Rev. B {\bf 81}, 241405R (2010).

\end{thebibliography}
\end{document}